\documentclass{article}
\usepackage{spconf,amsmath,graphicx}

\title{SAMO: Speaker Attractor Multi-Center One-Class Learning for Voice Anti-Spoofing }

\usepackage{url}
\usepackage{xcolor}

\usepackage{bm}
\usepackage{amssymb}
\usepackage{amsmath}
\usepackage{hyperref}
\usepackage{graphicx}
\usepackage{algpseudocode}
\usepackage{enumerate}
\usepackage{booktabs}
\usepackage{makecell}
\usepackage{threeparttable}
\usepackage{tabularx} 
\usepackage{multirow}
\usepackage{ragged2e}
\newcolumntype{L}{>{\RaggedRight\hangafter=1\hangindent=0em}X}
\usepackage[ruled, lined, linesnumbered, commentsnumbered, longend]{algorithm2e}

\begin{document}
\ninept

\twoauthors
 {Siwen Ding}
	{Columbia University, New York, NY, USA}
 {You Zhang, Zhiyao Duan\sthanks{This work is supported in part by a New York State Center of Excellence in Data Science award and synergistic activities funded by the National Science Foundation (NSF) under grant DGE-1922591. The authors would also like to thank Fei Jiang for the initial discussion of the multi-center idea.}}
	{University of Rochester, Rochester, NY, USA}




\maketitle

\begin{abstract}
Voice anti-spoofing systems are crucial auxiliaries for automatic speaker verification (ASV) systems. A major challenge is caused by unseen attacks empowered by advanced speech synthesis technologies.  Our previous research on one-class learning has improved the generalization ability to unseen attacks by compacting the bona fide speech in the embedding space. However, such compactness lacks consideration of the diversity of speakers. In this work, we propose speaker attractor multi-center one-class learning (SAMO), which clusters bona fide speech around a number of speaker attractors and pushes away spoofing attacks from all the attractors in a high-dimensional embedding space. For training, we propose an algorithm for the co-optimization of bona fide speech clustering and bona fide/spoof classification. For inference, we propose strategies to enable anti-spoofing for speakers without enrollment. Our proposed system outperforms existing state-of-the-art single systems with a relative improvement of 38\% on equal error rate (EER) on the ASVspoof2019 LA evaluation set.
\end{abstract}



\section{Introduction}
Automatic speaker verification (ASV) systems are essential for voice-based authentication that can recognize individuals through speech.
With the rise of advanced voice conversion and speech synthesis techniques, ASV technologies are susceptible to possible spoofing attacks, 
especially logical access (LA) attacks with synthetic speech 
~\cite{sahidullah2019introduction, nautsch2021asvspoof}. 
To tackle such vulnerability in ASV systems, a growing focus has been paid to spoofing countermeasure (i.e., CM or anti-spoofing) systems 
that often instrument deep learning models to differentiate speech spoken by humans (i.e., bona fide speech) and synthesized speech (i.e., spoofing attacks)~\cite{nautsch2021asvspoof}.

Existing research on speech anti-spoofing investigated different embedding extractions.
Recent end-to-end models have shifted from hand-crafted features such as LFCC~\cite{patel2015combining} and CQCC~\cite{todisco2016new, todisco2017constant} to raw waveforms, achieving state-of-the-art performance~\cite{tak2021end, hua2021towards,fu2022fastaudio,jung2022aasist}. 
Other efforts have focused on training strategies, such as data augmentation~\cite{tak2022rawboost, cohen2022study} and multi-task learning~\cite{mo2022multi}. Despite the impressive results, the main challenge remains on the generalization ability to unseen attacks, i.e., spoofing attacks generated by speech synthesis techniques that are not used in generating training data
\cite{sahidullah2019introduction}. While this is a pressing issue due to the rapid development of speech synthesis techniques, most cutting-edge anti-spoofing systems do not explicitly address this problem. 

\begin{figure}[t]
\centerline{\includegraphics[width=\columnwidth, height=0.9\textheight, keepaspectratio]{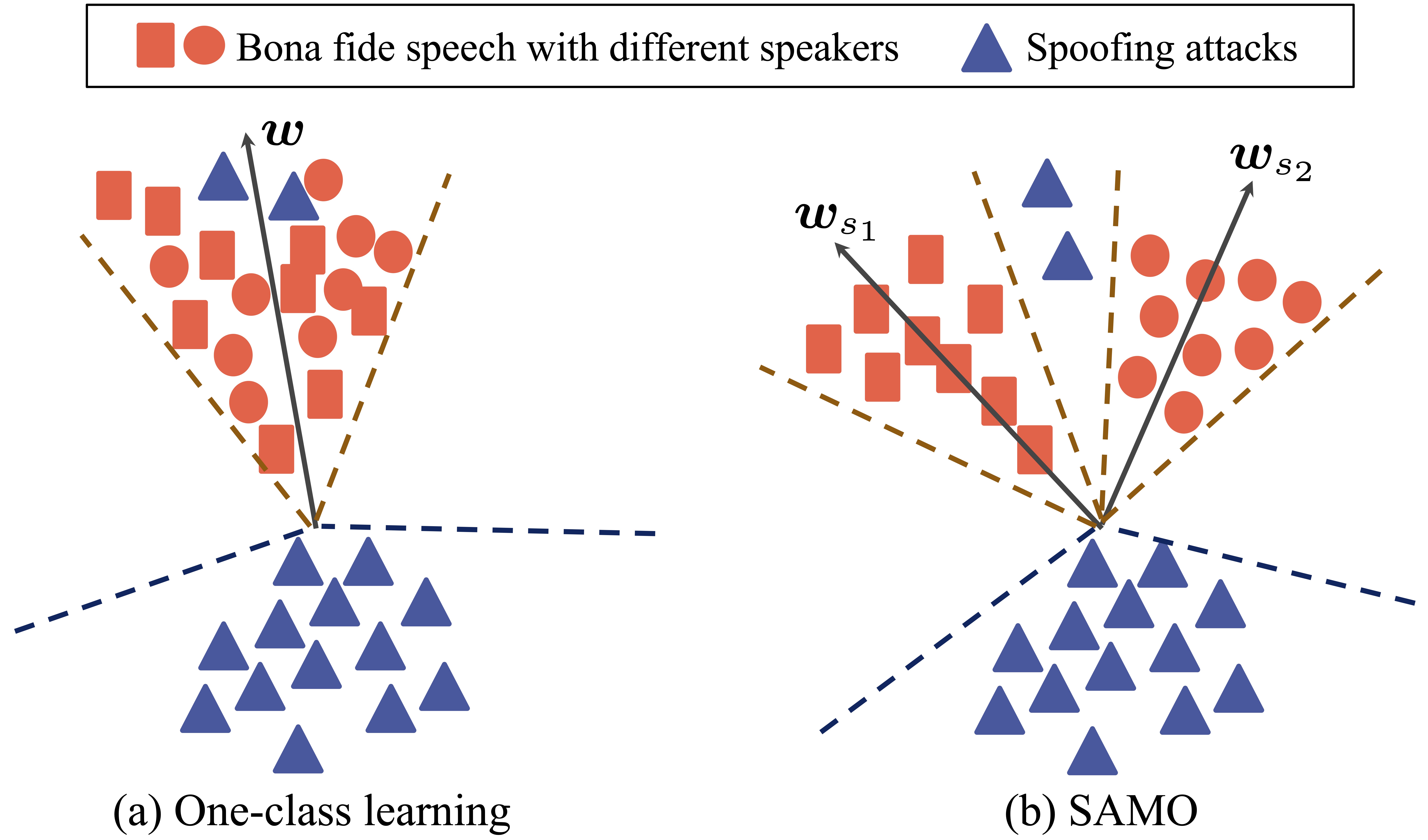}}
\caption{Comparison between one-class learning and our proposed SAMO method.}
\label{fig: samo}
\end{figure}

To address this generalization problem, in our prior work, we proposed the one-class learning approach~\cite{zhang21one} that tries to compact the bona fide speech into a cluster in the learned embedding space while pushing away spoofing attacks from this cluster. This approach has shown significant improvement in the generalization ability to unseen spoofing attacks~\cite{zhang21one}. 
However, due to the variety of timbre and speaking traits of different speakers, we have found that the bona fide speech of different speakers naturally forms multiple clusters in the embedding space. We argue that compacting different speakers into one cluster according to our original formulation~\cite{zhang21one} may have caused some misclassification of the spoofing attacks
~\cite{ghafoori2020deep}. 
For instance, as shown in Figure~\ref{fig: samo}(a), unseen spoofing attacks 
may lie in between multiple naturally separated clusters of bona fide speech, and compacting bona fide speech into a cluster is likely to coerce the spoofing attack samples into the cluster as well. It would make more sense to maintain the natural clustering of bona fide speech while pushing away spoofing attacks, as shown in (b).

In this work, we propose speaker attractor multi-center one-class learning (SAMO) to model speaker diversity while maintaining the generalization ability brought by one-class learning. We aim to find a speech representation for anti-spoofing that not only discriminates bona fide speech from spoofing attacks but also clusters bona fide speech according to speakers. After training, this model can be deployed to test speakers with or without enrollment utterances: With enrollment, the test utterance can be compared to the enrollment utterances for being classified into bona fide or spoofing attacks; Without enrollment, the test utterance can be compared to the learned attractors from training.
The proposed system achieves notable improvements over state-of-the-art single systems on the ASVspoof2019 LA dataset, suggesting the effectiveness of our proposed SAMO scheme. 

Our contributions are as follows: 1) We propose a novel multi-center one-class learning objective for anti-spoofing to respect the acoustic diversity of bona fide speech of different speakers, while discriminating bona fide and synthetic speech under the one-class learning paradigm. 2) We propose an algorithm with different optimization schedules to co-optimize the abovementioned aspects of the objective. 3) The proposed approach achieves the flexibility of deploying to speakers with or without enrollment utterances. Our work is reproducible with code at \url{https://github.com/sivannavis/samo}.

\section{Related Work}
\subsection{Speaker attributes in anti-spoofing}

Existing research explored the idea of leveraging speaker attributes in anti-spoofing. There are mainly three categories. Some investigated the person-of-interest approach, where in~\cite{castan22_odyssey}, a pre-trained anti-spoofing model is calibrated on bona fide and fake speech of the known identity. In~\cite{pian2022deepfake}, anti-spoofing is formulated as a speaker verification problem, and one-class strategies are employed to show promising in-the-wild generalization ability. However, they are limited to known identities whose bona fide speech is available during testing. 
Some used speaker embedding from speaker verification to assist anti-spoofing, either by concatenating the speaker embedding and the spoof embedding~\cite{chen2021spoofprint} or in a multi-objective learning fashion~\cite{mo2022multi, pan22_interspeech}. Some designed a joint system to integrate anti-spoofing with speaker verification, i.e., spoofing-aware speaker verification to optimize both objectives~\cite{jung2022sasv, zhang2022prob}. 

We propose a novel way to leverage the speaker attributes without requiring a reference set of bona fide speech for known identities in the test scenario.
Our method also has better simplicity with only one speech embedding extractor.

\subsection{Preliminary work on one-class learning}
\label{oc prem}
One-class classification is to address the classification scenario where the \textit{normal} class is well represented in the training set but the \textit{anomaly} class is not~\cite{khan2014one}.
With the distribution mismatch between training and test for spoofing attacks, specifically unseen synthetic attacks, the anti-spoofing problem fits the one-class classification setting.
The idea of one-class learning goes beyond one-class classification by jointly learning the embedding space and classifier. It was proposed in~\cite{zhang21one} to learn an embedding space for anti-spoofing using neural networks. It uses the One-Class-Softmax (OC-Softmax) loss function as the training objective:
\begin{equation}
\label{eq:oc-softmax}
\begin{aligned}
\mathcal{L}_\textit{OCS}=\frac{1}{N}\sum_{i=1}^N\log \big(1+e^{\alpha(m_{y_i}-\hat{\bm w}\hat{\bm x}_i)(-1)^{y_i}} \big),
\end{aligned}
\end{equation}
where $\hat {\bm x}_i \in \mathbb{R}^D$ represents the normalized speech embedding corresponding to the $i$-th utterance in a batch of $N$ samples, and $y_i \in \{0, 1\}$ is the label. $y_i = 0$ and $y_i = 1$ indicate that the utterance is bona fide and spoofing attack, respectively. $m_{y_i}$ refers to the boundary margins for the bona fide and spoof, respectively. $m_0 > m_1$.
Around the center vector $\hat {\bm w} \in \mathbb{R}^D$, $m_0$ compacts the distribution of the bona fide speech, while $m_1$ pushes away the spoofing attacks, and $\alpha$ is a scale factor. 

While this approach significantly improves the generalization ability to unseen attacks, it does
not consider the acoustic diversity of bona fide speech from different speakers, channels, and recording environments. Such diversity naturally forms multiple clusters of bona fide speech.
By squeezing bona fide speech embeddings into a single cluster, OC-Softmax faces the risk of misclassifying unseen spoofing attacks that would naturally lie in between those dense regions of bona fide speech.



\section{Methodology}

In this section, we describe our proposed method SAMO to address the abovementioned limitation of OC-Softmax~\cite{zhang21one}. The idea is to compact bona fide speech into multiple clusters instead of one, while pushing away spoofing attacks from all clusters. These clusters are formed based on speaker identity during training, and the cluster centers are called \textit{attractors}. During inference, we measure the 
cosine similarity between the test utterance and the nearest speaker attractor (if no enrollment data) or the target speaker's enrollment speech (if there is enrollment data) to classify it as bona fide or spoof. 

\subsection{Speaker attractors}
\label{attractor}
Extensive studies have manifested the intrinsic speaker diversity of bona fide speech in speech processing tasks, such as source separation~\cite{luo2018speaker, jiang2020speaker}, text-to-speech~\cite{goswami2022satts}, and speaker diarization~\cite{li2020speaker}.
This diversity naturally forms multiple clusters of bona fide speech in speech embedding spaces, where utterances from the same speaker tend to be close to each other. 
Leveraging this idea, we define a speaker attractor as a speaker-specific anchor in the embedding space to attract bona fide speech embeddings of the same speaker during training.
To compute and update such anchors,
we 
simply average the embeddings of each speaker's bona fide speech as a speaker attractor. 
Through training, the network weights and the embedding space are updated, and so do the attractors. However, the update of attractors does not happen in each iteration of the training; instead, it happens on a schedule explained in Section~\ref{algo}. 

Ideally, these attractors respect the natural clustering effect caused by the acoustic diversity of speakers while achieving good coverage of bona fide distribution for one-class learning. In the next section, we propose a loss function to achieve this goal.
After training, the learned attractors are fixed during inference. If the test speaker has no enrollment utterances, then the test utterance can be compared to these speaker attractors through the inner product to compute a score indicating whether it is bona fide or spoof. If the test speaker has enrollment utterances, then the test utterance can be compared to the enrollment utterances for the classification. The inference score is defined as
\begin{equation}
\label{eq:score1}
\begin{aligned}
S_\textit{SAMO} = \begin{cases}
   \hat{\bm w}_{s_i}\hat{\bm x}_i &\text{if $s_i$ is enrolled}  \\
   \underset{s}{\max}(\hat{\bm w}_{s}\hat{\bm x}_i), s \in \mathcal{S}_\textit{train}
    &\text{otherwise} 
\end{cases},
\end{aligned}
\end{equation}
where $\hat{\bm x}_i\in \mathbb{R}^{D}$ is the normalized speech embedding of the $i$-th test utterance, $s_i$ is the corresponding speaker, and $\hat{\bm w}_{s_i} \in \mathbb{R}^{D}$ is the average embedding of their enrolled utterances if any. $\hat{\bm w}_{s}$ are learned attractors of speaker $s$ in the training set. 

\subsection{Loss function for multi-center one-class learning}
\label{loss}
In Eq.~\eqref{eq:oc-softmax}, bona fide speech embeddings are compacted to a single cluster around the center $\hat{\bm w}$. Here we replace this single cluster with multiple clusters, each of which is around one attractor corresponding to one speaker in the training set, leading to the proposed SAMO loss function:
\begin{equation}
\label{eq:samo}
\begin{aligned}
\mathcal{L}_\textit{SAMO}=\frac{1}{N}\sum_{i=1}^N\log \big(1+e^{\alpha(m_{y_i}-d_i)(-1)^{y_i}} \big),
\end{aligned}
\end{equation}
where $d_i$ is calculated by
\begin{equation}
\label{eq:score2}
\begin{aligned}
d_i = \begin{cases}
   \bm{\hat{\bm w}_{s_i}\hat{\bm x}_i} &\text{if } y_i=0 \\
   \underset{s}{\max}(\hat{\bm w}_{s}\hat{\bm x}_i), s \in \mathcal{S_\textit{train}}
    &\text{if } y_i=1
\end{cases}.
\end{aligned}
\end{equation}
Here $\hat{\bm x}_i$ is the normalized speech embedding of the $i$-th utterance in the batch and $y_i$ is the corresponding label. $s_i$ is the corresponding speaker and $\hat{\bm w}_{s_i}$ is the normalized speaker attractor. $\alpha$ is the scale factor. $m_0$ and $m_1$ are the margins for the bona fide and spoof classes, respectively. 
$N$ is the number of utterances in the batch. 
$d_i$ is calculated as the cosine similarity to measure how similar the feature embeddings are to speaker attractors. As in Eq.~\eqref{eq:score2}, for a bona fide sample, the similarity is calculated between its normalized embedding $\hat{\bm x_i}$ and its speaker attractor $\hat{\bm w}_{s_i}$. Alternatively, when a sample belongs to the spoofing class, the similarity is between $\hat{\bm x_i}$ and its nearest speaker attractor among all training speakers $\mathcal{S}_\textit{train}$.
By minimizing this loss, the model can learn an embedding space that compacts speech utterances belonging to the same speaker and pushes away spoofing attacks from all speaker attractors.
This is different from OC-Softmax in Eq.~\eqref{eq:oc-softmax}, where all the training utterances are attracted to or pushed away from only one center $\hat{\bm{w}}$.

It is noted that the idea of multi-center has been employed in other tasks, such as face recognition~\cite{deng2020sub} and language recognition~\cite{ju2022masked} to better capture the local neighbor relationship of the embedding space. However, their centers are randomly assigned to samples of each class. Our proposed SAMO leverages the auxiliary speaker information in defining the clusters and only uses multiple centers for the bona fide class to enhance diversified compactness for one-class learning in the context of anti-spoofing.



\subsection{SAMO training algorithm}
\label{algo}
The proposed SAMO loss essentially has two sub-objectives: compacting the bona fide utterances spoken by the same speaker and pushing away spoofing utterances from all speaker attractors. Both sub-objectives use the attractors in the calculation.
By definition in Section~\ref{attractor}, each attractor is computed as the average embedding of the bona fide speech of each speaker, which changes as the embedding space changes. However, if the attractors are updated every time the embedding space updates (i.e., a mini-batch in network training), it may harm the convergence of the second sub-objective, i.e., pushing away spoofing utterances from all bona fide clusters. This is because any update of the attractors will change the optimization directions for the spoofing utterances. Empirically, we do find that frequently updating the attractors makes it difficult for the loss to converge.

We propose to update speaker attractors every $M$ epochs during training. The relative frequency of updates between the network weights (hence embedding space) and attractors is thus $M$ times the number of mini-batches in each epoch. In this way, the attractors serve as relatively stable anchors for attracting bona fide utterances and pushing away spoofed ones. In Section \ref{sec: exp}, we study how $M$ affects the system performance. At the beginning of training, we initialize speaker attractors as one-hot vectors in the embedding space, each representing the optimization direction of a speaker. The training process is described in Algorithm~\ref{alg: 1}.
\begin{algorithm}[t]
    \caption{SAMO Training Algorithm}
    \label{alg: 1}
    \SetKwInOut{KwIn}{Require}
    \SetKwInOut{KwOut}{Output}
    
    \KwIn{
    $T$: Total number of epochs \\
    $M$: speaker attractor update interval (\# epochs)
    }
    Initialize network $F$ with random weights\\
    Initialize speaker attractors $\bm w_s$ as one-hot vectors\\
    \For{$i \leftarrow 1$ \KwTo $T$}{
        \If{$i \operatorname{mod} M = 0$}{
                Update $\bm w_s$ as the average bona fide embedding for each speaker $s \in \mathcal{S}_\textit{train}$
        }
        Update $F$ by $\mathcal{L}_\textit{SAMO}$ with mini-batches \Comment{Eq.~\eqref{eq:samo}}
    }
    \KwRet{\normalfont{Optimized network} $F$ and \normalfont{speaker attractors} $\bm w_s$}
\end{algorithm}

\section{Experiments}
\label{sec: exp}

\subsection{Experimental setup}

\begin{table}[t]
\centering
\caption{Summary of the dataset used in our experiments, which is the target-only portion of the ASVspoof2019 LA corpus.}
\label{table: 1}
\resizebox{\columnwidth}{!}{
\begin{tabular}{@{}llllll@{}}
\toprule
\multirow{2}{*}{Partition} & \multirow{2}{*}{\begin{tabular}[c]{@{}c@{}}\# Speakers\end{tabular}} & \multirow{2}{*}{\# Enrollment} 
& Bona Fide     
& \multicolumn{2}{c}{Spoofing Attacks} \\ 
\cmidrule(l){4-6}
& &\hspace{1em}Utts & \# Utts & \# Utts   
& Attack Types     \\ 
\midrule
Train   & 20    & -  & 2580  & 22800 & A01$\sim$A06     \\
Dev & 10   & 142   & 1484  & 22296 & A01$\sim$A06     \\
Eval  & 48    & 696   & 5370  & 63882 & A07$\sim$A19     \\ 
\bottomrule
\end{tabular}}
\end{table}

As mentioned in Section~\ref{attractor}, the SAMO system, after training, can be used for test speakers with or without enrollment data. In our experiments, we test the systems in both scenarios.

\textbf{Dataset}.
We use utterances from ASVspoof2019 LA~\cite{wang2020asvspoof} as our data in all experiments. We use the same train/dev/eval splits as ASVspoof2019 LA. All attack types in the evaluation set are different from those in the training and development sets. Regarding speakers, we only keep the target speakers in the development and evaluation sets and discard the non-target speakers from these sets; 
The non-target speakers do not have enrollment data to support the scenario where enrollment data is used during inference, hence are adverse to our comparison between the two scenarios, both of which our method is able to deal with. Details of our dataset are shown in Table~\ref{table: 1}.

\textbf{Evaluation metrics}.
We use equal error rate (EER) and the minimum tandem detection cost function (min t-DCF) as the metrics for all experiments.
The EER metric is defined as the identical false acceptance rate and false rejection rate with a certain threshold of the inference score. 
Along with EER, min t-DCF evaluates all types of errors in speaker verification when the CM system is integrated into the "common" automatic speaker verification (ASV) system provided by the ASVspoof2019 challenge~\cite{todisco2019asvspoof}. As the ASV system is fixed, it can be viewed as a metric for the CM system when deployed to a speaker verification application~\cite{kinnunen2018t}. 

\textbf{Comparison methods}.
We compare SAMO with two state-of-the-art anti-spoofing methods: AASIST~\cite{jung2022aasist} and OC-Softmax~\cite{zhang21one}. AASIST is a model based on graph neural networks with integrated spectro-temporal attention. It uses the softmax loss in training. We train AASIST with its reported configurations in~\cite{jung2022aasist}.
In~\cite{zhang21one}, OC-Softmax uses a backbone network based on deep residual network ResNet-18 with one-class learning loss mentioned in Section~\ref{oc prem}. To remove the effect of backbone network architecture, we change the network backbone to AASIST but maintain the OC-Softmax loss function, using $\alpha=20, m_0=0.5$, and $m_1=-0.2$. 

\textbf{Implementation details}.
For all experiments, we use the architecture of AASIST~\cite{jung2022aasist} as our speech embedding network. 
A 160-d embedding vector is computed from the last hidden layer. 
For our proposed SAMO, we set the update interval $M = 3$; the scale factor $\alpha = 20$; the margins $m_0 = 0.7$, and $m_1=0$.
We train all systems on a single GPU with Adam optimizer \cite{kingma2014adam} and a learning rate of $10^{-4}$ with cosine annealing learning rate decay for 100 epochs and save the model in each epoch. We choose the model that achieves the lowest EER on the development set in the scenario where enrollment data is used.

\subsection{Results and discussions}


\begin{table}[t]
\centering
\caption{Comparison of our proposed SAMO with Softmax and OC-Softmax on the target-only portion of the ASVspoof2019 LA evaluation set. All the systems use AASIST~\cite{jung2022aasist} backbone. The average (best) results across 3 trials with random training seeds are shown.}

\label{table:2}
\small
\begin{tabular}{@{}lll@{}}
\toprule
Method 
&EER(\%) &min t-DCF \\ 
\midrule
Softmax &1.74 (1.25) & 0.0583 (0.0425)\\
OC-Softmax &1.25 (1.17) & 0.0415 (0.0393)\\

SAMO (test w/o enrollment)   &1.09 (0.91)   &0.0363 (0.0306)\\
SAMO (test w/ enrollment)
& \bf{1.08 (0.88)}    &\bf{0.0356 (0.0291)}\\
\bottomrule
\end{tabular}
\end{table}


\textbf{Comparison with state-of-the-art methods.} 
Table~\ref{table:2} shows the EER and min t-DCF comparisons between SAMO and the two comparison methods on the target-only portion of the ASVSpoof2019 LA evaluation set. We report the average results with 3 different random seeds for all models and report the best results in parentheses. We can see that:
1) The results of our proposed method SAMO with two inference strategies both demonstrate advantages over other comparison methods with notable margins, e.g., SAMO with enrollment decreases EER of
Softmax and OC-Softmax by relative 38\% (1.08\% vs 1.74\%) and 14\% (1.08\% vs 1.25 \%), respectively.
2) Compared to Softmax, OC-Softmax improves the system performance by 28\% relatively, which agrees with the advantages of one-class learning over the binary classification reported in~\cite{zhang21one}. 
3) Compared to OC-Softmax, SAMO further improves the performance by 14\% relatively, indicating the advantage brought by the multi-center modeling of bona fide speech in SAMO. 


\begin{table}[t]
\centering
\caption{Ablation experiments for SAMO. Results of test scenarios without and with enrollment data are both presented.}
\label{table:3}
\begin{threeparttable}
\resizebox{\columnwidth}{!}{
\begin{tabular}{@{}llll@{}}
\toprule
\multirow{2}{*}{Setup} & \multirow{2}{*}{Configuration} & \multicolumn{2}{c}{Test w/o enroll (w/ enroll)}\\
& &EER(\%) &min t-DCF\\ 
\midrule
1&SAMO   & \bf{1.09 (1.08)}    &\bf{0.0363 (0.0356)}      \\
2   &one-hot and fixed attractors & 47.93 (49.50) & 0.9999 (0.9980)\\
3   &w/o speaker attractor update &1.54 (1.55) &0.0504 (0.0503)\\
4   &update every epoch ($M$=1)  &1.33 (1.33)   &0.0442 (0.0437)\\
5   &update every 10 epochs ($M$=10)  &2.36 (2.77)   &0.0792 (0.0868)\\
\bottomrule
\end{tabular}}
\end{threeparttable}
\end{table}

\textbf{Ablation studies.} Table~\ref{table:3} investigates the effects of important components and parameters of our proposed SAMO system. We report average results for all setups with 3 random seeds.
To see the effects of leveraging bona fide speech data in speaker attractors, Setup 2 initializes the speaker attractors as one-hot vectors in the embedding space without further updates during training.
Results show that this setup does not learn a meaningful embedding space. 
Setup 3 initializes the attractors as one-hot vectors first, then only updates attractors once at the beginning of the second epoch. Compared with Setup 2, this setup shows that simply including bona fide speaker information in attractors without any training schedule helps with learning a useful embedding space. Compared with SAMO, it also shows that a proper update schedule is essential for achieving desirable results. 
To see the effects of our training algorithm in terms of the speaker attractor update interval $M$ (\# epochs) in Algorithm~\ref{alg: 1}, we vary its value in Setups 4 and 5. Their results show that speaker attractor update interval has a crucial impact on the anti-spoofing performance in both test scenarios. 
Compared to SAMO ($M=3$), Setup 4 ($M=1$) updates the speaker attractors more frequently, and the model might not have optimized the embedding space well enough before an attractor update, 
while Setup 5 ($M=10$) updates them less frequently, and the model might not have updated the attractors adequately for facilitating the embedding space learning. 

\begin{figure}[t]
\centerline{\includegraphics[width=1.05\columnwidth, height=\textheight, keepaspectratio]
{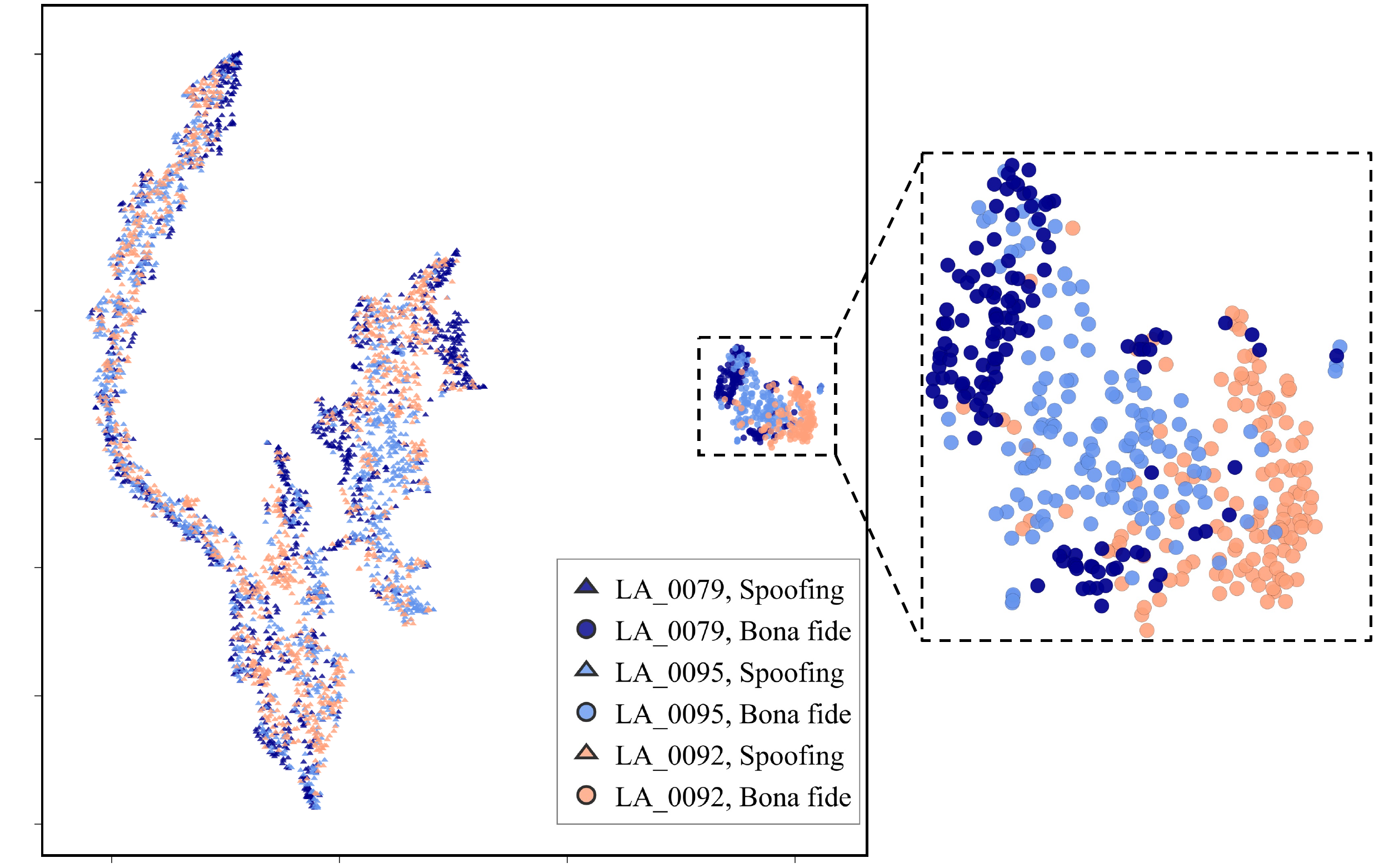}}
\caption{2D t-SNE visualization of SAMO feature embeddings of bona fide and spoofed speech of three speakers with speaker ID LA\_0079, LA\_0092, and LA\_0095 in the training partition. }
\label{fig: viz}
\end{figure}

\textbf{Embedding visualizations.}
To comprehend the effects of speaker attractors in the embedding space, we pick three speakers in the training set and transform their 160-D embedding vectors to 2-D in Figure~\ref{fig: viz} with t-SNE~\cite{van2008visualizing}.
From the visualization, we can see that the bona fide utterances are grouped in a small region at the right side of the figure, while the spoofing attacks are pushed away from this region, occupying a large area on the left. This shows that SAMO follows the general one-class learning framework. Within the bona fide utterance region, we can also see that utterances of the three speakers are generally clustered according to speaker identity. This shows that SAMO models the bona fide class as speaker clusters. Both observations show that the SAMO algorithm performs as expected as the objective function Eq.~\eqref{eq:samo}. 

The current training set only contains 20 speakers. We believe that with a larger variety of speakers in the training set, the benefit of SAMO could be demonstrated even more since the speaker attractors will better represent the bona fide embedding space.




\section{Conclusion}
We proposed speaker attractor multi-center one-class learning (SAMO) as a novel representation learning framework for voice anti-spoofing. It models speaker diversity in bona fide speech to build compact speaker-specific clusters in the embedding space. The approach is to compact bonafide speech around speaker attractors and isolate spoofing attacks from all attractors.
Experiments show that SAMO improves anti-spoofing performance over state-of-the-art methods, including AASIST and OC-Softmax.
For future work, we plan to extend the SAMO idea to model other speech attributes, 
such as device~\cite{zhang21ea_interspeech} and codec variations~\cite{liu2022asvspoof}.



\vfill\pagebreak

\footnotesize
\bibliographystyle{IEEEbib}
\bibliography{main}

\end{document}